\documentclass{edm_article}
\usepackage{xcolor}

\begin{document}

\title{An Approach to Detect Abnormal Submissions for CodeWorkout Dataset}

% Submissions for EDM are double-blind: please do not include any author names or affiliations in the submission. 
% Anonymous authors:
\numberofauthors{5}
% \author{
% Anonymous\\
%        \affaddr{Anonymous Institution}\\
%        \email{anonymous@anonymous.edu}
% }
%An example of how to include
% multiple authors is below for after the paper has been accepted.


% \numberofauthors{8} %  in this sample file, there are a *total*
% % of EIGHT authors. SIX appear on the 'first-page' (for formatting
% % reasons) and the remaining two appear in the \additionalauthors section.
% %
\author{
% % You can go ahead and credit any number of authors here,
% % e.g. one 'row of three' or two rows (consisting of one row of three
% % and a second row of one, two or three).
% %
% % The command \alignauthor (no curly braces needed) should
% % precede each author name, affiliation/snail-mail address and
% % e-mail address. Additionally, tag each line of
% % affiliation/address with \affaddr, and tag the
% % e-mail address with \email.
% %
% % 1st. author
\alignauthor
Alex Hicks\\
       \affaddr{Dept of Computer Science}\\
       \affaddr{Virginia Tech}\\
       \affaddr{Blacksburg, VA}\\
       % \email{alexhicks@vt.edu}
% 2nd. author
\alignauthor
Yang Shi\\
      \affaddr{Dept of Computer Science}\\
       \affaddr{Utah State University}\\
       \affaddr{Logan, UT}\\
       % \email{yang.shi@usu.edu}
% % 3rd. author
\alignauthor 
Arun-Balajiee Lekshmi-Narayanan\\
       \affaddr{Intelligent Systems Program}\\
       \affaddr{University of Pittsburgh}\\
       \affaddr{Pittsburgh, PA}\\
       % \email{arl122@pitt.edu}
\and  % use '\and' if you need 'another row' of author names
% % 4th. author
\alignauthor Wei Yan\\
       \affaddr{School of Informatics, Computing and Cyber Systems}\\
       \affaddr{North Arizona University}\\
       \affaddr{Flagstaff, AZ}\\
       % \email{wei.yan@nau.edu}
% % 5th. author
\alignauthor Samiha Marwan\\
       \affaddr{Dept. of Computer Science}\\
       \affaddr{University of Virginia}\\
       \affaddr{Charlottesville, VA}\\
       % \email{samihamarwan21@gmail.com}
% % 6th. author
% \alignauthor Charles Palmer\\
%        \affaddr{Palmer Research Laboratories}\\
%        \affaddr{8600 Datapoint Drive}\\
%        \affaddr{San Antonio, Texas 78229}\\
%        \email{cpalmer@prl.com}
}
% % There's nothing stopping you putting the seventh, eighth, etc.
% % author on the opening page (as the 'third row') but we ask,
% % for aesthetic reasons that you place these 'additional authors'
% % in the \additional authors block, viz.
% \additionalauthors{Additional authors: John Smith (The Th{\o}rv{\"a}ld Group,
% email: {\texttt{jsmith@affiliation.org}}) and Julius P.~Kumquat
% (The Kumquat Consortium, email: {\texttt{jpkumquat@consortium.net}}).}
% \date{30 July 1999}
% Just remember to make sure that the TOTAL number of authors
% is the number that will appear on the first page PLUS the
% number that will appear in the \additionalauthors section.

\maketitle

%  did a relatively fine pass in the abstract and introduction, and also left some general comments in the other sections -- please let me know if additional help is needed from my end -- in short, there are several important things in my opinion:
% 1) I wanted to change the framing to "anomaly detection" instead of cheat detection -- anomaly could be anomaly for us but cheating is like detecting something we don't have ground truths.
% 2) This is a description paper, so we only need to talk about the method we introduce and give a short and sweet comparison to MOSS. We will probably not have enough room to address the research question of how this would actually make an impact in student modeling, though this is an important motivation in the introduction and discussions.
% 3) Importantly, we will need to address how good our results are -- no need to say this is excellent, but still make sure to spend 1 paragraph on result interpretations.
% Good luck; generally, it's in shape and fits well into the target track in CSEDM. There are just some more steps to go. I would like to rely on Arun/Alex to address the concerns, and Samiha/Wei to give it final passes if time permits. Cheers!

\begin{abstract}
% Cheating in practice programming assignments has become very common due to the rise of technology and the pressure to succeed. 
% Student interactions can be logged by interfaces CodeWorkOut (CWO) where the log data can be used to provide personalizable problem recommendations with incrementally challenging exercises depending on the level of the student knowledge. However, anomalies in the student interactions, such as cheating to solve programming puzzles, could introduce a hidden bias in the log data, so the students will not be benefited by using the system. Classical methods such as MOSS can be used to detect code plagiarism. However, it is limited to cheating detection and cannot further detect other abnormal events such as student gaming a system with multiple attempts of similar solutions to a particular programming puzzle. This description paper focuses on our preliminary study to analyzing log data with anomalies. The goal of our work is to overcome the abnormal instances when modeling personalizable recommendations to students learning programming.

%%%% Samiha's abstract recommendation%%%%%%
Students’ interactions while solving problems in learning environments (i.e. log data) are often used to support students’ learning. For example, researchers use log data to develop systems that can provide students with personalized problem recommendations based on their knowledge level. However, anomalies in the students’ log data, such as cheating to solve programming problems, could introduce a hidden bias in the log data. As a result, these systems may provide inaccurate problem recommendations, and therefore, defeat their purpose. Classical cheating detection methods, such as MOSS, can be used to detect code plagiarism. However, these methods cannot detect other abnormal events such as a student gaming a system with multiple attempts of similar solutions to a particular programming problem. This paper presents a  preliminary study to analyze log data with anomalies. The goal of our work is to overcome the abnormal instances when modeling personalizable recommendations in programming learning environments.

% We investigated students' potential cheating instances in a publicly available dataset. We found that using MOSS to detect cheating within the dataset may not work, especially for short programming assignments.  Finally, we discuss some existing approaches similar to the techniques that we used, to tackle the challenge of unsupervised cheating detection in data, thereby offering solutions to clean data before being used for student modeling.
\end{abstract}

 % Replace with your own 3-5 keywords

\section{Introduction}

% Such flawed instances are referred to as abnormal instances, which could be cheating or other behaviors unrelated to learning. While these instances could lead to studies on students' abnormal behaviors such as cheating, gaming the system and distractions. The other instance that is more interesting to consider, where they do not spend the effort or time to learn the concepts to solve the programming quiz,  when the goal is to estimate students' knowledge and learning.
% In these cases, the interaction logs that carry the response provided by the student as "their" original response, are in fact, not being true to the knowledge level of the student. This could, hence, be incorrectly picked up by metrics used in normalized student models as the student gaining knowledge, when, in fact, they might not have picked up the required concepts from solving the problem. .

% 2) If not, can further approaches be developed to identify such instances? 
% 2) What potential benefits could the detection bring for tasks of student modeling?  

% From our discussions, we present a possibly novel technique that could be used in the context of student interaction log data collected from web--based practice programming platforms for the purposes of student modeling and other kinds of predictive modeling.

% \section{Background \& Related Work}

% \subsection{General Approaches to Curb Cheating}
{Students cheating to submit programming solutions is a common occurrence. Cheating can be of any kind -- copying solutions to the problem available online, by other students learning programming with the course or by other means of plagiarism.} Generally, researchers have explored methods to curb cheating in the context of academic integrity~\cite{allen2023systemic}. Some techniques that could work~\cite{karnalim2022educating} include the detection of collusion and continual feedback to students to encourage them towards better academic integrity. There is a tendency for students to cheat when solving programming puzzles or practice assignments. When online log data is collected using the interaction logs of the interfaces for programming assignments, there is a risk for some of these anomalies to be recorded among regular student interaction logs. This could potentially affect student modeling approaches that use the interaction logs to make recommendations for students~\cite{brusilovsky2007user}.

% This, in turn, benefits in the development of astudent model that tracks student learning and predict student performance in an introductory programming class. 

Student modeling in the context of solving programming assignments like the Normalized Student Modeling for Programming~\cite{carter2015normalized} use Error Quotient and Watwin score that measure changes help estimate student knowledge or understanding~\cite{carter2015normalized,price2020progsnap2}.  In other cases, student modeling facilitates the identification and prediction of students’ learning profiles in tutoring systems, which, in turn, enables such systems to be adaptive and personalized to students’ needs~\cite{umer2023current}. This makes them sensitive to the quality of the data and anomalies created by students gaming the system or cheating / plagiarizing solutions may cause the model to \textit{overestimate} or \textit{underestimate} student knowledge or understanding of the introductory programming concepts.

% However, many student modeling techniques are highly sensitive to data, relying on clean data for accurate predictions, a condition that is often not met with publicly shared data. 

For example, a study conducted by Hellas et al. found instances where students copied content to complete their assignments \cite{hellasplag2017}. This behavior can significantly compromise the quality of student modeling approaches applied to these data. Moreover, these cheating instances may lead to erroneous predictions, revealing a threat to the field of student modeling technology. 

{Another example discussed by Sosnovsky and colleagues~\cite{sosnovsky2018detection} discusses student modeling anamolies observable as sudden changes in the learning rate of a student when learning with an adaptive educational system. This could be attributed to any form of assistance offered to the student by a more experienced or knowledgeable peer indicated Low-High-Low or High-Low-High patterns in the student's learning rate.}

% Several student modeling techniques have been applied to students' interaction log data to measure and support student learning while solving problems in programming learning environments. Several popular metrics have been used to evaluate changes in a code written by a student as a response to a programming assignments. 

% These metrics utilize information from student interaction logs, which are used by normalized student models for programming~\cite{carter2015normalized}. This implies that if the metrics measure \textit{flawed} instances of student logs then they may tend to \textit{overestimate} or \textit{underestimate} a student's interaction with the system. 

% As a result, the use of these flawed instances will affect the student model that maps the student behavior to their interactions to measure student long-term learning. In addition, the use of these flawed instances may affect the predictions of students' knowledge gain and performance in the course. 

To address this challenge, researchers have developed tools for detecting plagiarism in students' code (e.g., \cite{bowyer1999experience}). One of the most popular approaches is “The Measure Of Software Similarity (MOSS)”, an open-source tool designed to identify similarities between students' programming assignments \cite{bowyer1999experience}.  However, to our knowledge, there is no evidence that researchers apply cheating detection methods on online shared data before applying log data analysis and student modeling. 

{We present a work in progress, where we look into this aspect closely in order to mitigate anamolies in student submissions : 1) using classical methods like Measure of Software Similarity (MOSS), 2) alternative approaches of analyzing log data). We use the CodeWorkout (CWO) programming dataset (as introduced in \cite{shi2023kc})\footnote{ https://pslcdatashop.web.cmu.edu/DatasetInfo?datasetId=3458}. While the use of generative AI has been very popular now, this dataset was collected before 2021 when Generative AI was not generally used to cheat when submitting programming solutions}. 

\begin{table*}[]
    \centering
    \begin{tabular}{|l|r|r|r|r|r|r|l|}
    \hline
     & 
      {avg\_score} &
      {median\_score} &
      {first\_score} &
      {last\_score} &
      {n\_attempts} &
      {one\_shot} &
      {condition} 
    \\ \hline
     &
      &
       &
      &
       &
       &
       &
       \\
    X-Grade (before) &
      -0.209966 &
      -0.187436 &
      -0.287831 &
      0.062299 &
      0.220599 &
      -0.359111 &
      Unclean \\
       &
      &
       &
      &
       &
       &
       &
       \\
    X-Grade (after) &
      0.326563 &
      0.440488 &
      0.457784 &
      0.333629 &
      0.376384 &
      0.234074 &
      Clean \\  
      &
      &
       &
      &
       &
       &
       &
       \\
       \hline
    \end{tabular}
    \caption{Comparison between correlation data across potential indicators with final course grade before and after cleaning suspicious submissions.}
    \label{tab:corr}
    
\end{table*}

\section{Methods \& Analysis}

% \subsection{Dataset}

% For our analysis, we consider datasets that consist of log data on student program problem-solving, that have been made publicly available as part of the CSEDM Data Challenge~\footnote{https://sites.google.com/ncsu.edu/csedm-dc-2021/dataset}. We use the CodeWorkout (CWO) dataset that consists of interaction and submission log data for the Fall 2019 programming assignment submissions made by students in ProgSnap2 format~\cite{price2020progsnap2}. 

In this work, we compare two ways to analyze abnormal submissions:

    % \item \textbf{Detecting Cheating with Log Data Analysis:} We only analyze the student submission trace data without actual code to detect student cheating behavior.

\textbf{Proposed method: Log Data Analysis.} We used two main identifiers to explore anomalies such as suspected cheating behaviors from submission log data: the number of submission attempts before completing the exercise, and the elapsed time between correct submissions. The choice of these variables correlates with the possibility that students who attempt and submit a correct solution on their first attempt could be cheating. We discuss more details on this below.

\textbf{Baseline method: MOSS.} {MOSS is a tool used to detect cheating in programming submissions. The tool works by taking into all the students' submissions and comparing them pairwise for similarities}. We compared students' code submissions using MOSS to identify similarities in submissions for a selected set of problems from a collection of easy, medium, and hard assignments made available on CWO.

\section{Results \& Discussions}

% as shown in the Figure~\ref{fig:MOSS-hard}.
% ,~\ref{fig:MOSS-medium} and ~\ref{fig:MOSS-easy}.

% \begin{figure}[h!]
%     \centering
%     \includegraphics{images/MOSS-hard.png}
%     \caption{We observe that MOSS seems to detect the two solutions to be somewhat different with its color coding, but the 3 chunks of the code are almost similar. Additionally it shows around only 80\% similarity}
%     \label{fig:MOSS-hard}
% \end{figure}

% \begin{figure}
%     \centering
%     \includegraphics[width=\textwidth]{images/MOSS-medium.png}
%     \caption{}
%     \label{fig:MOSS-medium}
% \end{figure}

% \begin{figure*}
%     \centering
%     \includegraphics[width=\textwidth]{images/MOSS-easy.png}
%     \caption{Since the code is small for certain submissions MOSS may list them as the same}
%     \label{fig:MOSS-easy}
% \end{figure*}
\subsection{MOSS Detection Results}

\begin{figure*}[h!]
    \centering
    \includegraphics[width=\textwidth]{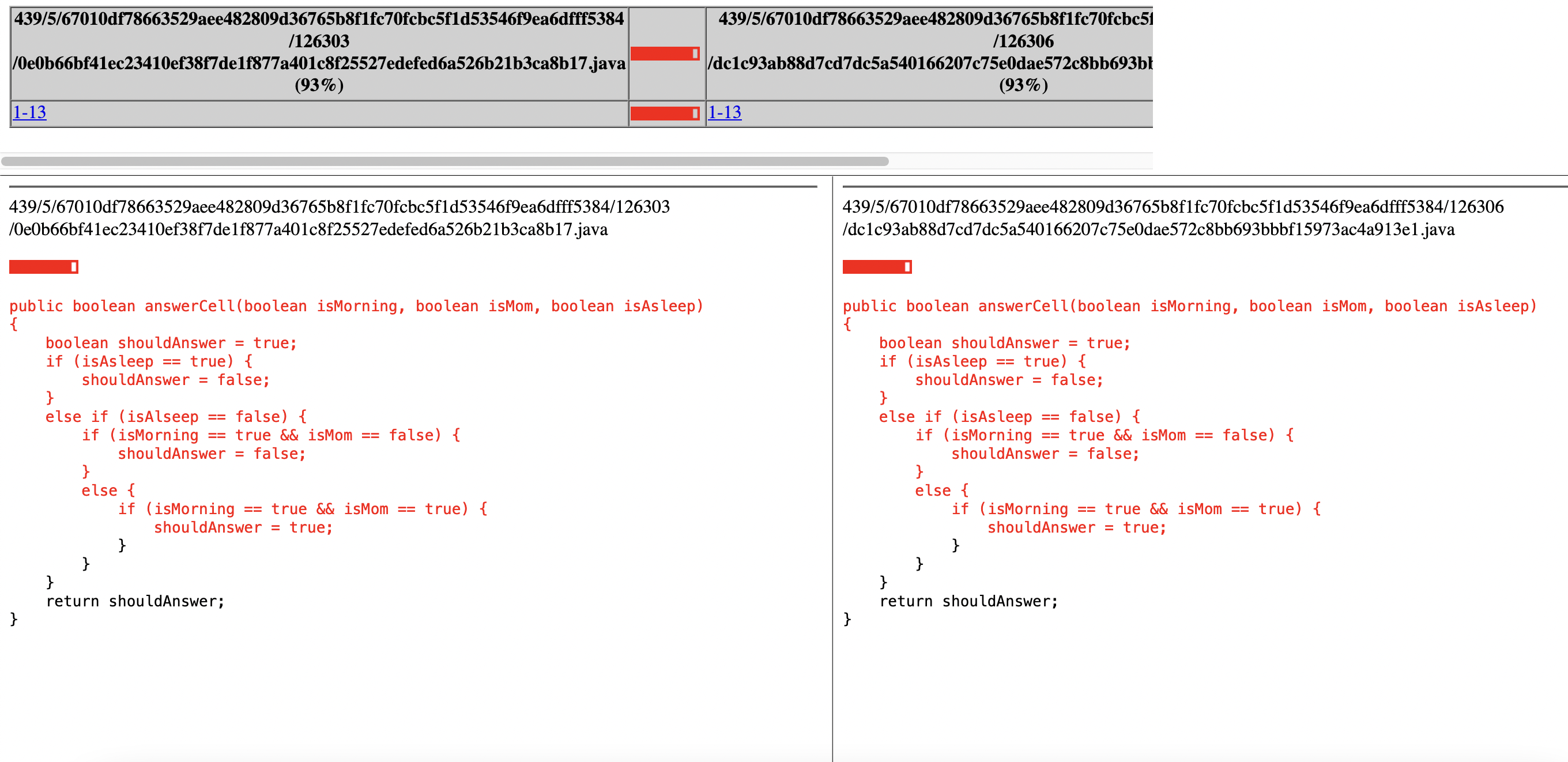}
    \caption{MOSS results for comparing two CWO submissions}
    \label{fig:moss_easy}
\end{figure*}

% In order to validate our initial findings, we attempted to replicate our results using MOSS to identify similar submissions across students. 
We further evaluated whether accessible and common cheating detection tools such as MOSS can be applied to detect students' cheating in this dataset. However, we found that running MOSS across CWO exercises led to high rates of similarity on a majority of students' submissions. In addition, we found no clear difference between students whom we previously identified and those whom we believe that have engaged authentically with the CWO exercises. We hypothesize that this failure could be due to the size of the solutions to several CWO exercises. Some solutions to these exercises could be just 10 lines of source code as these problems are well-constrained and target specific learning goals. Hence, these problems may not have possible alternative solutions (refer Figure~\ref{fig:moss_easy}). Students like those in the example may end with 93\% of their solutions matching despite no indications of anomalous behaviour. This indicates that identifying an acceptable threshold for MOSS detection on CWO exercises is unreasonable and highlights the need for other options.

\subsection{Log Data Analysis Detection}

\begin{figure*}[h!]
    \centering
    \includegraphics[width=\textwidth]{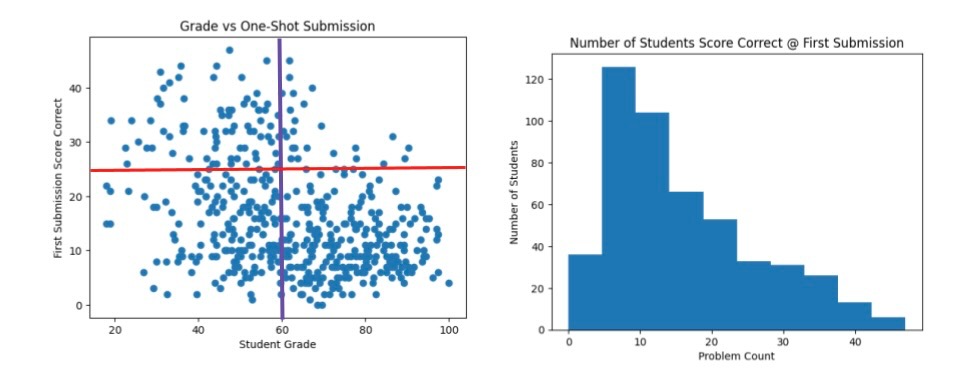}
    \caption{Cluster plot of student first submission scores on exercises (y-axis) and their final grades(x-axis) (left); Histogram of the number of students (y-axis) with the number of exercises they achieved correctly (x-axis) on first submissions.}
    \label{fig:detect_cheating}
\end{figure*}

% We conducted an exploratory data analysis on the attributes of the ProgSnap2 data~\cite{price2020progsnap2} that would allow us to identify this cheating behavior implicitly. From this analysis, we identified some difficult-to-understand correlations, such as the negative correlation between the score on a student's first attempt at a given problem and their final course grade. To further investigate, we created a potential measure that calculates the number of students who achieve a fully correct solution on their first attempt.
We calculate students' ``one shot'' percent, or the percent of CWO exercises where a student correctly answers an exercise on their first attempt. In Table \ref{tab:corr}, this is represented as the one\_shot column and is calculated as a correlation with the student's final course grade. Once this value was calculated, we were able to compare the differences between the correlations on a student's first score on a given problem to how often they were getting their first attempt fully correct and found a suspicious difference. Figures~\ref{fig:detect_cheating} shows the relationship between the first scores of the students' submission to the exercises and their final exam scores, and the figure on the right shows the distribution of the first scores of the students' submission. While many students perform well on their first submissions of exercises, showing their mastery of programming skills, only a small subset match this performance in the course as a whole. Specifically, students who perform well in the CWO exercises on their first attempt, often do not perform well for their final grade of the course. This preliminary data analysis did not make intuitive sense and led us to further investigate this phenomenon using more traditional methods, including MOSS.

\section{Limitations and Future Work}

This preliminary investigation focused only on CWO submissions, but we hope this data cleaning approach can be generalized to other datasets that use the ProgSnap2 format. We also hope to continue investigating the metadata about submissions included in this format to find more accurate indicators of cheating behavior in the programming snapshot data. {While MOSS is generally used to compare final students' submissions with other final students' submissions, in future work, we will consider the case for running MOSS with sequential data where submissions made on platforms like CWO that allow multiple submissions. For example, we could compare attempt 1 of a student  1 with attempt 2 of  student 2 and so on to see if a students copy each others' solutions from their first attempt onwards or after trying multiple attempts, failing and then cheat to proceed to the next programming problem on the CWO platform.}

%ACKNOWLEDGMENTS are optional
\section{Acknowledgments}
We thank the contributions by Dr. Thomas Price for his guidance on this work. We also thank the 2023 Session of LearnLab Summer School Organizers and our sponsors Dr. Peter Brusilovsky and SPLICE project PI(s) for bringing us all together to work on this.

\bibliographystyle{abbrv}
\bibliography{sigproc}  % sigproc.bib is the name of the Bibliography in this case
%\balancecolumns
% 
% That's all folks!
\end{document}